\documentclass[aps,prb,showpacs,superscriptaddress,groupedaddress,twocolumn]{revtex4-1}
\usepackage{amssymb}
\usepackage{amsmath}
\usepackage{amsfonts}
\usepackage{graphicx}
\usepackage{latexsym}
\usepackage{ulem}
\usepackage{color}
\usepackage{subfigure}

\expandafter\let\csname equation*\endcsname\relax
\expandafter\let\csname endequation*\endcsname\relax

\usepackage[verbose,hypertexnames=false,bookmarksopenlevel=1,filecolor=blue,
linkcolor=blue,citecolor=blue,pdfstartview=FitH,bookmarksopen,bookmarksnumbered,
colorlinks,plainpages=false,linktocpage]{hyperref}

\begin{document}
%--------------------------------------------------------------------------------
\title{Lifetimes of Magnons in Two-Dimensional Diluted Ferromagnetic Systems}
%--------------------------------------------------------------------------------
\author{Akash Chakraborty}
\author{Paul Wenk}
\author{John Schliemann}
\affiliation{
Institut f\"ur Theoretische Physik, Univerist\"at Regensburg, 93040 Regensburg, Germany
}
\date{\today}
%--------------------------------------------------------------------------------
\begin{abstract}
  Spin dynamics in low dimensional magnetic systems has been of
  fundamental importance for a long time and has currently received an
  impetus owing to the emerging field of nanoelectronics.  Knowledge
  of the spin wave lifetimes, in particular, can be favorable for
  future potential applications. We investigate the low-temperature
  spin wave excitations in two-dimensional disordered ferromagnetic
  systems, with a particular focus on the long wavelength magnon
  lifetimes. A semi-analytical Green's functions based approach is
  used to determine the dynamical spectral functions, for different
  magnetic impurity concentrations, from which the intrinsic linewidth
  is extracted. We obtain an unambiguous $q^4$ scaling of the magnon
  linewidth which is ascribed to the disorder induced damping of the
  spin waves, thereby settling a longstanding unresolved issue on the
  wave-vector dependence. Our findings are also in good agreement with
  previous theoretical studies on Heisenberg
  ferromagnets. Additionally, we demonstrate the futility of using the
  low moments associated with the spectral densities to evaluate the
  magnon dispersions and lifetimes. 
\end{abstract}
%--------------------------------------------------------------------------------
\pacs{75.30.Ds, 73.21.-b, 75.50.Pp}
%--------------------------------------------------------------------------------
%--------------------------------------------------------------------------------
\maketitle
\section{INTRODUCTION}
\label{sect:Introduction}
%--------------------------------------------------------------------------------
\ Two-dimensional magnetic systems have been a subject of intensive
investigation for almost half a century now. Both ferro- and
anti-ferromagnetic systems have been studied extensively,
experimentally as well as theoretically, revealing a myriad of
interesting properties including the discovery of high-temperature
superconductivity in doped two-dimensional
cuprates\cite{bednorz1986}. One of the important aspects of these
systems which has continued to attract interest is the magnetic excitations 
which are of fundamental relevance to understand the spin dynamics. Knowledge 
of the collective spin wave excitations can provide
valuable insight into their dynamical response as well as their
thermodynamic behavior. The advancement of experimental techniques,
such as ferromagnetic resonance spectroscopy and inelastic neutron
scattering, have been of immense help in exploring this field
meticulously\cite{coldea2001,lumsden2009}. Inelastic neutron scattering is one of the most powerful
and versatile tools as the long wavelength spin excitations, better known as
magnons, can be probed directly and accurately. One of the most
thoroughly studied systems, in this context, is the two-dimensional
Heisenberg anti-ferromagnet
Rb$_2$Mn$_{1-x}$Mg$_x$F$_4$\cite{cowley1977,birgeneau1977,birgeneau1980}, which 
was investigated by means of neutron scattering techniques to study the
low-temperature magnetic excitations, viz.\ the magnon dispersion,
linewidths, and lineshapes, as well as the critical exponents near the
transition temperature. This led to similar studies on
K$_2$Cu$_{1-x}$Zn$_x$F$_4$\cite{wagner1978}, which is a quasi
bi-dimensional ferromagnet. In the aforementioned studies, good
agreement with numerical calculations, available at that time, was
also reported. However, despite the existence of innumerable studies,
one important feature which has eluded understanding, over the
decades, is the wave-vector dependence of the magnon lifetime
(inversely proportional to the linewidth), especially in the long
wavelength limit ($q\rightarrow 0$).
%--------------------------------------------------------------------------------

\ Spin waves in Heisenberg ferromagnets, in the low-energy limit, were
studied theoretically as early as the sixties by
Murray\cite{murray1966}. The author calculated the spin wave energies
and the scattering cross section, within the Born approximation, and
reported a $q^5$ scaling of the magnon lifetime. The exchange
interactions, in this case, were restricted to nearest neighbors
only. Similar $q^5$ dependence was also found in amorphous Heisenberg
ferromagnets, in the low-temperature and long wavelength limit, by
using an effective medium approximation\cite{singh1978}. In this case,
however, spatially dependent extended couplings were assumed between
the magnetic sites. Based on Green's functions calculations,
Mano\cite{mano1982} also predicted an identical behavior of the
lifetime in the long wavelength limit. The finite linewidth of the
excitations, which increased rapidly with decreasing wavelength, was
attributed to the randomness in the magnitude of the spins. Also the
discrepancy between the observed magnetization behavior and that
predicted by elementary spin wave theory was believed to originate
from this finite linewidth of the spin waves. On the contrary, similar
spin wave studies in amorphous ferromagnets by
Kaneyoshi\cite{kaneyoshi1978} led to a $q^7$ dependence of the
linewidth. This was an outcome of using a quasi-crystalline
approximation, which is essentially a virtual-crystal-like
approach. Within this approximation, the magnon dispersion reduces to
that of an ideal crystal, wherein the disorder effects are completely
neglected. In yet another study, based on the two-magnon interaction
theory of Heisenberg ferromagnets, a leading order $q^2$ scaling of
the magnon lifetime was suggested by Ishikawa \textit{et 
al.}\cite{ishikawa1981}. However, in most of the aforesaid studies,
the systems under consideration were three-dimensional Heisenberg
ferromagnets and there was no clear mention of the dimensional
dependence. It was only later that Christou and
Stinchcombe\cite{christou1986} investigated the low-temperature spin
excitations in bond-diluted Heisenberg ferromagnets from a more
generalized perspective. Using a diagrammatic perturbation theory, the
authors obtained a $q^{d+2}$ ($d>1$, is the dimensionality) scaling of
the magnon linewidth. Although the discussion was extended to the more
relevant site-diluted systems, the exchange interactions were again
restricted to nearest neighbors only.
%--------------------------------------------------------------------------------

\ Thus, the lack of a general accord on the issue of linewidth scaling
becomes apparent from the widely varying predictions available in the
literature. Moreover, considerable attention and interest have also
been devoted to the case of anti-ferromagnets, including even
lately\cite{bayrakci2013}. In a very recent study\cite{akash2015}, on three-dimensional 
disordered ferromagnets, a $q^5$ scaling of the magnon linewidth, in the long wavelength limit, 
was reported using similar  numerical approaches as implemented here.
This served as a further motivation
behind the current study of the magnetic excitations in two-dimensional 
ferromagnetic systems, with a view to identify the dimensional dependence of 
the scaling of the magnon lifetimes.
Also a proper knowledge of the lifetimes is
not only of fundamental interest but can also be of practical
importance, as we shall discuss later. In this article, we provide a
comprehensive and detailed analysis of the low-temperature spin wave
excitations in two-dimensional site-diluted ferromagnets, in the
presence of extended exchange interactions. The calculations have been
performed on sufficiently large system sizes and a proper statistical
sampling over disorder is also taken into account. We lay special
emphasis on the correct evaluation of the magnon linewidths in the
long wavelength limit. In the process, we demonstrate that determining
the correct wave-vector dependence of the lifetimes constitutes a
non-trivial task. In addition, we also discuss the nature of the
magnon density of states, the spectral functions, as well as the
magnon dispersion over a relatively broad concentration range.
%--------------------------------------------------------------------------------
\section{HEISENBERG MODEL AND EXCHANGE COUPLINGS}
\label{sect:Model}
%--------------------------------------------------------------------------------
\ We start with the Hamiltonian describing $N_{imp}$ interacting spins
(${\bf S}_{i}$) randomly distributed on a square lattice of $N$ sites, given
by the dilute Heisenberg model
\begin{align}
H=-\sum_{i,j} J_{ij}p_{i}p_{j} {\bf S}_{i}\cdot{\bf S}_{j}
\label{Hamiltonian}
\end{align}
where the sum $i,j$ runs over all sites and the random variable
$p_i$=1 if the site is occupied by an impurity or otherwise zero. We
consider classical spins ($\mid$${\bf S}_{i}$$\mid=S$) on a square
lattice, with lattice spacing $a$, and with periodic boundary
conditions. The distribution of the spins, in this case, is completely random and uncorrelated; in other 
words the probability of a spin to be placed at site $i$ is independent of the neighboring sites. This 
is in contrast to a previous study\cite{akash2014} on the magnetic excitations in inhomogeneous diluted systems, 
where well-defined spherical clusters of spins were considered.
Spin-orbit coupling is neglected as this would lead to anisotropy in the system 
which is not the primary focus here. The effects of spin-orbit on the magnon lifetimes in 
two-dimensional systems were studied in\cite{zakeri2012}.
All calculations, in the present work, are performed at
$T$=0 K. The concentration of magnetic impurities in the system is
denoted by $x$ ($=N_{imp}/N$). The Hamiltonian,
Eq.~(\ref{Hamiltonian}), is treated within the self-consistent local
random phase approximation (SC-LRPA), which is essentially a
semi-analytical approach based on (finite temperature) Green's
functions. Within this approach, the retarded Green's functions are
defined as
\begin{align}
G^c_{ij}(\omega)=\int_{-\infty}^{\infty}G^c_{ij}(t)e^{i\omega t}dt
\label{retarded_GF}
\end{align}
where $G^c_{ij}(t)$=$-i\theta(t)\langle[{\bf S}_i^+(t),{\bf
  S}_j^-(0)]\rangle$, describe the transverse spin fluctuations, and
$\langle\ldots\rangle$ denotes the expectation value, and `$c$' the
disorder configuration index. After performing the Tyablikov
decoupling\cite{tahirkheli1962,tyablikov1967,nolting2009} (assuming
magnetization along the $z$-axis) of the higher-order Green's
functions which appear in the equation of motion of
$G^c_{ij}(\omega)$, we obtain
\begin{align}
(\omega{\bf I}-{\bf H}_{eff}^c){\bf G}^{c}={\bf D}
\label{eff_matrix}
\end{align}
where ${\bf H}_{eff}^c$, ${\bf G}^{c}$, and $\bf D$ are
$N_{imp}\times N_{imp}$ matrices. The effective Hamiltonian matrix elements
are 
\begin{align}
  ({{\bf H}_{eff}^c)}_{ij}= -\langle{S_i^z}\rangle
  J_{ij}+\delta_{ij}\sum_{l}\langle{S_l^z}\rangle J_{lj}
\end{align}
and the diagonal matrix
\begin{align}
  D_{ij}= 2\langle S_i^z\rangle \delta_{ij}.
\end{align}
For a given temperature and disorder configuration, the local
magnetizations $\langle{S_i^z}\rangle$ ($i=1,2,\ldots,N_{imp}$)
have to be calculated self-consistently. However, since we are
interested at $T$=0 K, where the ground state is assumed to be fully
polarized, all $\langle{S_i^z}\rangle$ are equal to $S$ in this
case. We shall not go into further details of the method here, as the
accuracy and reliability of the SC-LRPA to handle disorder (dilution)
effects in different contexts have been discussed and established on numerous
previous occasions (for details see Refs.~\onlinecite{georges2005,georgesprb2005,sato2010}).
The virtual crystal approximation, as a possible alternative approach, fails in these systems 
as will be discussed in Sec.~\ref{sect:Lifetime}.
%--------------------------------------------------------------------------------

\ The exchange interactions are assumed to be of the form
$J_{ij}$=$J_{0} r_{ij}^{-\alpha}$, where $r_{ij}=|{\bf r}_i-{\bf
  r}_j|$. In most of the previous studies the exchange couplings were
restricted to nearest neighbors only, but in realistic systems these
interactions extend well beyond the nearest neighbors. Moreover, the
choice of the couplings is motivated by the theoretical proposition
put forward in Ref.~\onlinecite{bruno2001}, wherein the author extends
the Mermin-Wagner theorem\cite{mermin1966} to Heisenberg and $XY$
systems with long-range interactions. It is stated that a
$d$-dimensional system ($d$=1 or 2) with monotonically decaying
interactions as $\mid$$J_{\textbf r}$$\mid \propto r^{-\alpha}$ cannot
have ferro- or anti-ferromagnetic long-range order at $T> 0$, if
$\alpha \ge 2d$. For RKKY-like interactions (long-range oscillatory
nature) magnetic order could be strictly ruled out for the
one-dimensional systems, but only for certain cases in the
two-dimensional ones. It was later proved by Loss, Pedrocchi, and
Leggett\cite{loss2011}, again as an extension of the Mermin-Wagner
theorem, that no long-range magnetic order is possible in one or
two-dimensional systems at a finite temperature, in the presence of
RKKY interactions. The choice of exponentially decaying couplings can
also be ruled out in this case as they satisfy the Mermin-Wagner
theorem trivially. Hence, this led us to the choice of the exponent
$\alpha$=3 for the couplings, which implies that in our
two-dimensional systems long-range (ferro-)magnetic order at a finite
temperature is not excluded by the Mermin-Wagner theorem.
Also, since the couplings are isotropic and all ferromagnetic ($J_{ij}>0$)
there is no frustration expected and hence the collinear state can be safely assumed to 
be the ground state. This in turn leads to only positive eigenvalues in the 
magnon spectrum which will become clear in the following calculations of the magnon DOS.
%--------------------------------------------------------------------------------
\section{MAGNON DENSITY OF STATES AND SPECTRAL FUNCTION}
\label{sect:DOS}
%--------------------------------------------------------------------------------
\begin{figure}[htbp]\centerline
{\includegraphics[width=\columnwidth,angle=0]{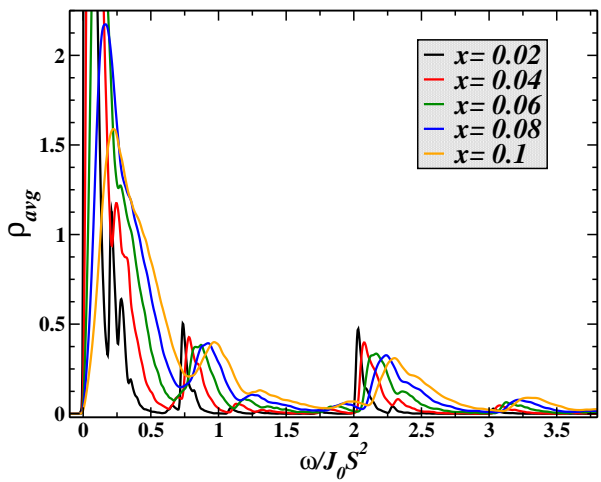}}
\caption{(Color online) Average magnon DOS $\rho_{\text{avg}}$ as a function of energy 
$\omega$ plotted for different concentrations $x$.}
\label{fig1} 
\end{figure}
\begin{figure}[htbp]
\centering
\subfigure[]{
  \includegraphics[width=0.45\columnwidth,angle=0]{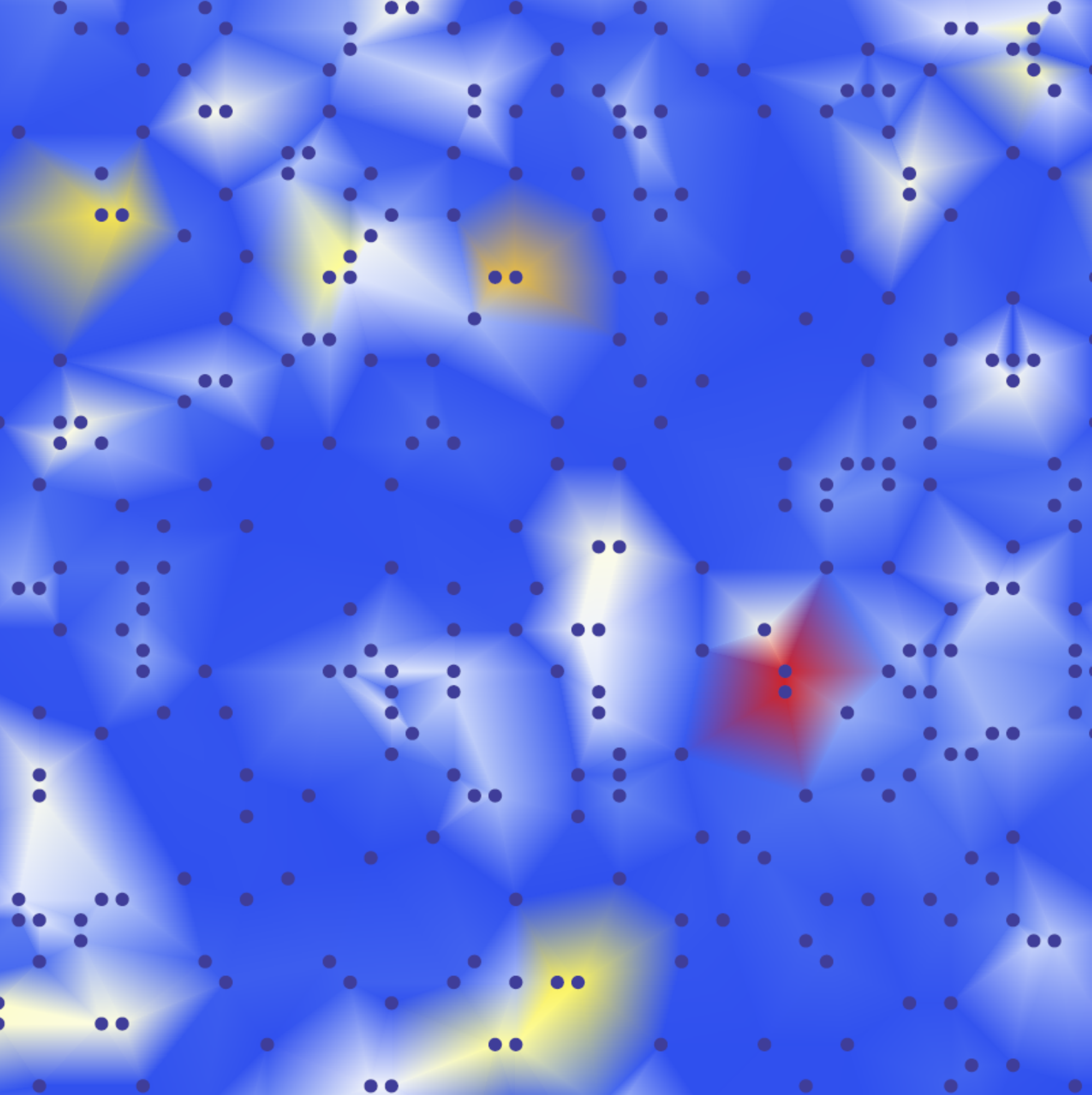}
\label{fig2a}
}
\subfigure[]{
  \includegraphics[width=0.45\columnwidth,angle=0]{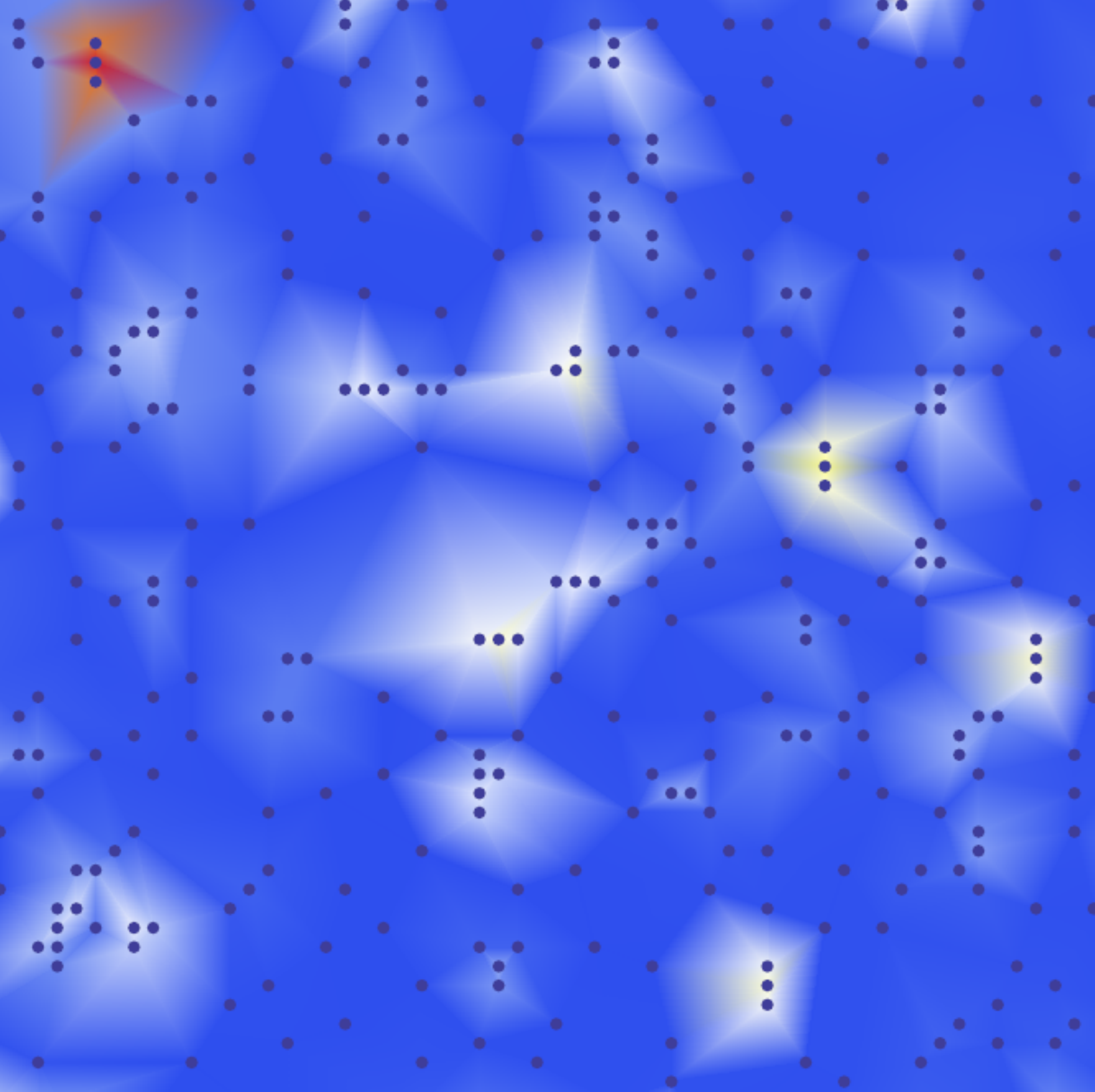}
\label{fig2b}
}
\subfigure{
  \includegraphics[width=0.5\columnwidth,angle=0]{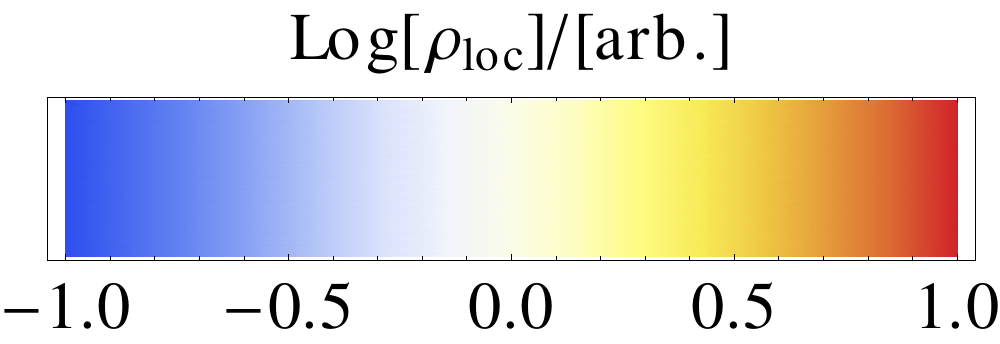}
\label{fig2c}
}
\caption[Optional caption for list of figures]{(Color online) Distribution of the
  local magnon DOS at energies (a) $\omega/(J_0S^2)=2.2$, and (b)
  $\omega/(J_0S^2)=3.2$, for an impurity concentration of
  $x=0.1$. Shown is a part of the lattice of size $L$=$200a$ in
  coordinate space. The dots indicate the positions of the spins ${\bf
    S}_i$.}
\label{fig2}
\end{figure}

 
%--------------------------------------------------------------------------------
\ From the retarded Green's functions defined above one can calculate
the average magnon density of states (DOS), which is given by
$\rho_{\text{avg}}(\omega)=(1/N_{\text{imp}})\sum_i\rho_i(\omega)$,
where $\rho_i(\omega)=-1/(2\pi S)\Im[ G_{ii}(\omega)]$ is the local
magnon DOS\@. Fig.~\ref{fig1} shows the average magnon DOS as a
function of the energy for different impurity concentrations. The DOS
have been averaged over a hundred disorder configurations, although it
was found that typically 25 configurations were sufficient for each
impurity concentration. We observe irregular features in the DOS which
become more pronounced with increase in dilution. On decreasing the
concentration from $x=0.1$ to $x=0.02$, a significant increase in
weight around the low energy end of the spectrum is observed.  This
increase in weight is attributed to the increase in the fraction of
impurities which are weakly connected to the rest. These isolated
impurity regions have their own zero-energy modes which in turn
contribute to the DOS at the low energies. In order to gain a better
insight into this behavior we look at the distribution of the local
DOS shown in Fig.~\ref{fig2a} and ~\ref{fig2b}, at two different
energies 2.2 $J_0S^2$ and 3.2 $J_0S^2$, respectively for $x=0.1$. Here
we can clearly identify certain impurity regions, of typically two or
three impurities, which are weakly coupled to the surrounding
impurities. These can be seen to make a higher contribution to the
DOS. (For more details see Fig.~\ref{fig7}, App.). Note that the
distribution shown corresponds only to a part of the lattice from a
$200a \times 200a$ system. With increasing dilution the average
separation between the spins increases and hence the effective
coupling decreases. This accounts for the increase in the irregular
features observed in the DOS for $x=0.02$.
%--------------------------------------------------------------------------------
%--------------------------------------------------------------------------------

%--------------------------------------------------------------------------------
\begin{figure*}[htbp]
  \centering
  \includegraphics[width=0.75\textwidth,angle=0]{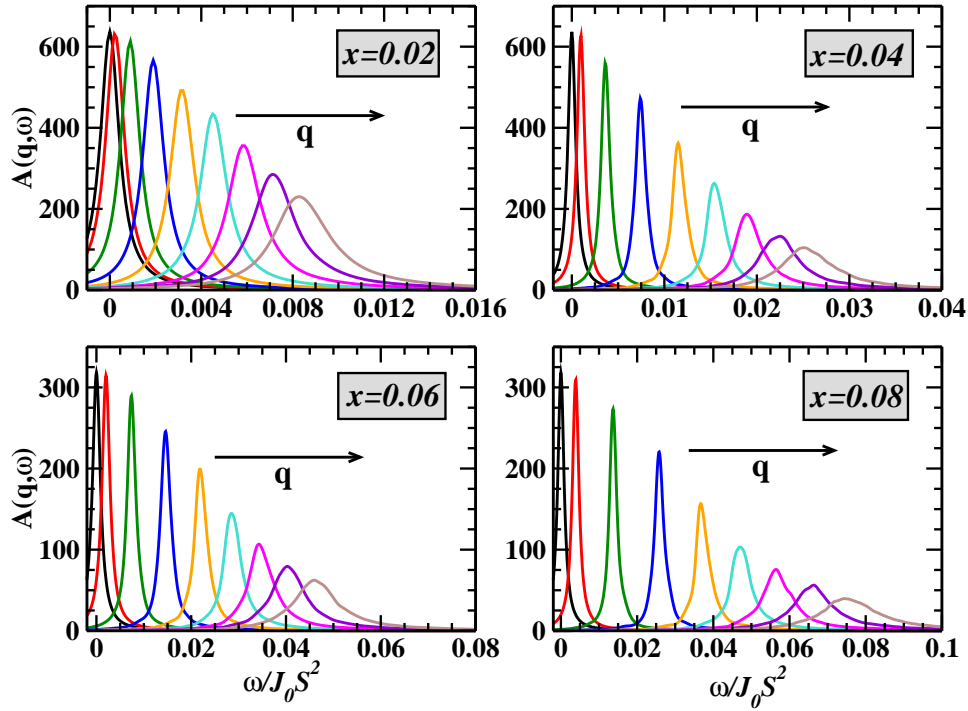}
  \caption{(Color online) Average spectral function $A({\bf q},\omega)$ as a function
    of the energy with ${\bf q}$=$n(2\pi/La)\{1,0\}^\top$ (where
    $n\in\mathbb{N}$), corresponding to four different concentrations
    $x$. The system size is $L^{2}=300a\times300a$.}
  \label{fig3}
\end{figure*}
%--------------------------------------------------------------------------------
\ The dynamical spectral function, also known as the structure factor,
provides valuable insight into the underlying spin dynamics of a
system. Experimentally this can be probed by inelastic neutron
scattering and ferromagnetic resonance to a good accuracy. The
averaged spectral function is defined by
\begin{align} 
{A}({\bf q},\omega) :={}& -\left\langle \frac{1}{2\pi S}
\Im[G^{c}({\bf q},\omega)]\right\rangle_c,
\end{align}
where $G^{c}({\bf q},\omega)$ is the Fourier transform of the retarded
Green's function $G^{c}_{ij}(t)$, and $\langle\dots\rangle_{c}$
denotes the configuration average. Fig.~\ref{fig3} shows the averaged
spectral functions as a function of energy for four different
concentrations. The $A({\bf q},\omega)$'s are averaged over a few
hundred disorder configurations, and the results are plotted only in
the [1 0] direction of the Brillouin zone, for progressively
increasing momentum $\bf q$, since the focus is on the long wavelength
regime here. It should be noted that the [0 1] direction is equivalent
to the [1 0] direction in this case, due to the lattice symmetry. Also
note that for $q\gg 2\pi/(La)$ the deviation from rotation invariance
is not negligible. Well-defined excitations are found to exist only
for small values of $\bf q$, in each case. For increasing $\bf q$, the
excitation peaks become broader and develop a tail extending toward
the higher energies. On decreasing the concentration from 0.08 to 0.02
the zone of stability of the well-defined magnon modes is found to
decrease by almost one order of magnitude. Also the excitations become
increasingly asymmetric with increase in the momentum. This increase
in asymmetry is associated with the crossover from propagating
low-energy spin waves to localized or quasi-localized excitations
(fractons)\cite{orbach1987} at higher energies. Here, the term
\textit{localized} implies that the excitations are quite broad in
energy at fixed wave-vectors, or rather quantitatively the excitation
energy is much larger than the linewidth (i.e.\ the full-width at
half-maximum).

\ The nature of the spectral functions is similar to what was observed
by neutron scattering experiments in
Mn$_x$Zn$_{1-x}$F$_2$\cite{uemura1986,uemura1987}, which is a
three-dimensional randomly diluted anti-ferromagnet. The authors
measured sharp spin waves near the zone center which broadened
progressively with the wave vector approaching the
zone-boundary. These findings were attributed to a crossover from
low-energy extended spin waves (magnons) to localized high-energy
excitations (fractons), which was further consistent with the
theoretical conjecture of fractons in disordered percolating
networks\cite{orbach1987,aharony1985}. A recent numerical
study\cite{mucciolo2004} on site-diluted two-dimensional
anti-ferromagnets also reveal the existence of localized excitations
at high energies. The authors evaluated the inverse participation
ratio (IPR), for different dilutions and different system sizes in
order to establish the nature (extended or localized) of the states,
although the largest system size studied was only $32a \times
32a$. These studies provide relevance and also additional motivation
to study the two-dimensional ferromagnets from this aspect. The proper
and accurate evaluation of the spectral functions, as we shall see in
what follows, constitutes a vital task since the magnon dispersion as
well as the lifetime can be directly extracted from them.

%--------------------------------------------------------------------------------
\section{MOMENTS ANALYSIS}
\label{sect:Moments}
%--------------------------------------------------------------------------------
\begin{figure}[htbp]\centerline
{\includegraphics[width=\columnwidth,angle=0]{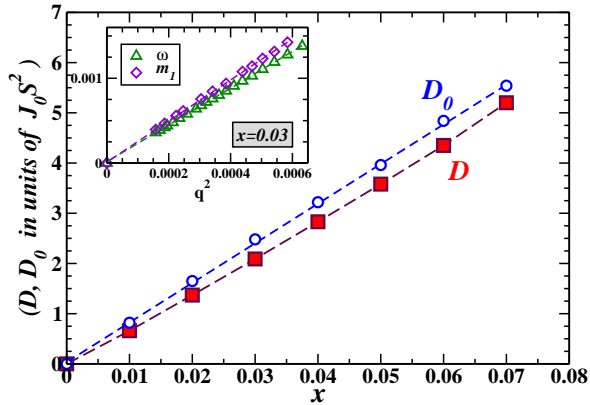}}
\caption{(Color online) Spin stiffness $D$ and effective spin stiffness $D_{0}$ as a
  function of $x$. ($m_{1}(q)\approx D_{0}q^2$, where $m_1$ is the
  first moment associated to the spectral density). The inset shows a
  comparison of the excitation energies extracted from $A({\bf
    q},\omega)$ and the first moments $m_1$, respectively, for the
  case of $x=0.03$.}
\label{fig4} 
\end{figure}
%--------------------------------------------------------------------------------
\ Before embarking into further details of the long wavelength magnon
properties, we define the moments associated with the spectral
density. The $n$-th moment is defined by
\begin{align}
m_n({\bf q})=\int_{0}^{\infty} \omega^{n} A({\bf q},\omega) d\omega
\label{moments}
\end{align}
In the limit $q\rightarrow 0$, it can be shown that $m_{1}({\bf q})
\approx D_{0}q^{2}$\cite{georges2007}, where we call $D_{0}$ as the
effective spin wave stiffness. It is also well known that in the long
wavelength limit the dispersion in ferromagnetic systems is quadratic
in $q$, $\omega({\bf q}) \approx Dq^{2}$, where $D$ denotes the spin
stiffness coefficient. The moments, as sometimes found in the
literature\cite{motome2002,motome2005}, are used in the spectral
function analyses as a good approximation to estimate the excitation
energy and linewidth, especially in the presence of
disorder. Nonetheless, the accuracy and the viability of this
assumption is subject to further examination. In order to address
this, as a first step, we numerically calculated the dispersion from
the first moment and then compared it to the \textit{real} excitation
energy $\omega({\bf q})$ extracted from the $A({\bf q},\omega)$ peaks
shown in Fig.~\ref{fig3}. The results for the particular case of
$x=0.03$ are plotted in the inset of Fig.~\ref{fig4}. As can be seen,
in the small $q$ limit, both $m_{1}(q)$ and $\omega({\bf q})$ are
linear in $q^2$ but the first moment fairly overestimates the real
magnon energies. This is better reflected when we extract the
respective spin stiffness coefficients, $D_{0}$ from $m_{1}(q)$ and
$D$ from $\omega({\bf q})$, and plot them against the concentration as
shown in Fig.~\ref{fig4}. For all considered $x$, the effective spin
stiffness is larger than the actual spin stiffness, overestimating by
15-20\% in each case. This clearly demonstrates that the first moment
is not a reliable quantity to evaluate the spin stiffness in these
diluted systems as it fails to reproduce the magnon energies
precisely.

%--------------------------------------------------------------------------------
\begin{figure}[htbp]\centerline
{\includegraphics[width=\columnwidth,angle=0]{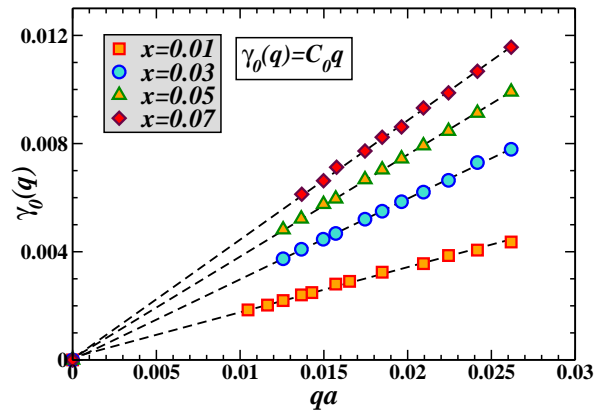}}
\caption{(Color online) Effective linewidth $\gamma_{0}$ (in units of $J_{0}S^{2}$)
  as a function of $q$ (in the [1 0] direction), for different
  concentrations $x$, Eq.~(\ref{eff_linewidth}). The dashed lines
  indicate the linear fits.}
\label{fig5} 
\end{figure}
%--------------------------------------------------------------------------------
\ The other relevant quantity of interest is the intrinsic linewidth
of the magnetic excitations. The linewidth gives a measure of the
excitations' broadening due to disorder, which maybe magnetic or
structural disorder, or due to the magnon-magnon interactions. One can
obtain the linewidth from the second moment from the
relation\cite{motome2002}
\begin{align}
\gamma_{0}({\bf q})= \sqrt{m_{2}({\bf q})-m_{1}^{2}({\bf q})}
\label{eff_linewidth}
\end{align}
where $\gamma_{0}({\bf q})$ is the effective linewidth. In
Fig.~\ref{fig5} we have plotted this effective linewidth as a function
of {\bf q} for four different impurity concentrations. We find that in
the small-$q$ limit the linewidth is linear in $q$ for all considered
$x$. The same holds true for all other intermediate concentrations,
which are not shown here. Consequently, we end up with $\omega \propto
q^2$ and $\gamma_{0} \propto q$, in the limit $q\rightarrow 0$. This
indicates that the magnetic excitations are incoherent or localized
around the $\Gamma$-point, since $\gamma_{0} > \omega$. However, this
is somehow contrary to what we have observed in the spectral functions
shown in Fig.~\ref{fig3}, where the excitations are well defined for
small $q$ values. Hence, we can safely assume that the effective
linewidth obtained from the moments does not correspond to the real
linewidth of the excitations. The same discrepancy was also
demonstrated for the case of Ga$_{1-x}$Mn$_x$As\cite{georges2007}, a
well-known III-V diluted ferromagnetic semiconductor, where the
lattice has an fcc structure. Note that similar linear $q$-dependence,
obtained from the moments analysis, was reported by the authors in
disordered double-exchange systems\cite{motome2005}. Determining the
correct $q$-dependence of the intrinsic linewidth, in the long
wavelength limit, requires further detailed analysis which is
elucidated in the following.
%--------------------------------------------------------------------------------
\section{SCALING OF MAGNON LIFETIME}
\label{sect:Lifetime}
\begin{figure}[htbp]
\centering
\subfigure{
\includegraphics[width=\columnwidth]{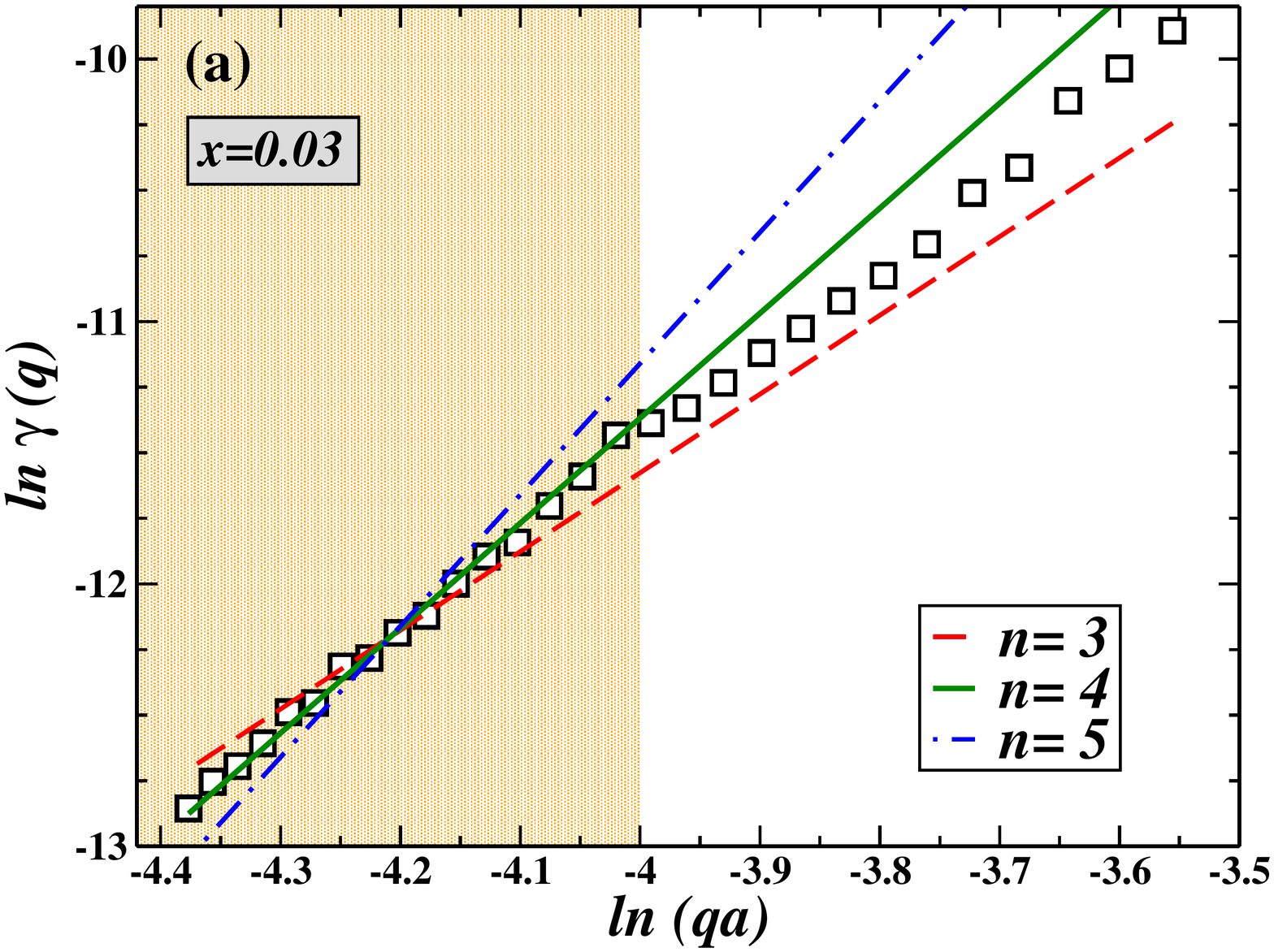}
\label{fig6a}
}
\subfigure{
\includegraphics[width=\columnwidth]{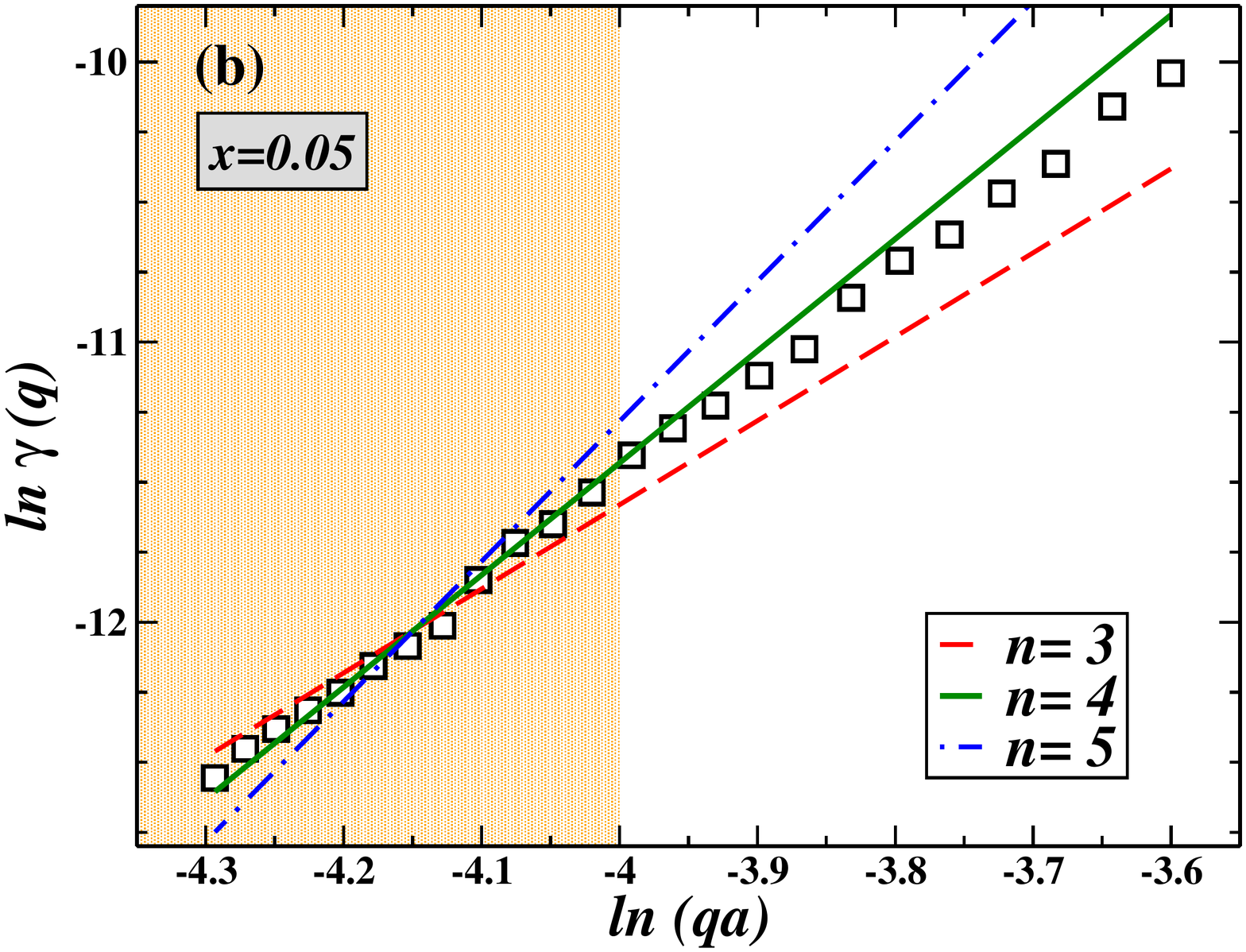}
\label{fig6b}
}
\label{fig6}
\caption[Optional caption for list of figures]{(Color online) Logarithm of the
  magnon linewidth $\gamma$ (in units of $J_{0}S^{2}$) as a function
  of ln $(qa)$, for (a) $x=0.03$, and (b) $x=0.05$. Dashed (red),
  solid (green), and dot-dashed (blue) lines indicate linear fits of
  the form $n\ln(qa)+C$, ($n=3$, 4, and 5), for the linewidth data
  within the shaded region.}
\end{figure}

%--------------------------------------------------------------------------------
\ We extract the linewidth, which is the full-width at half-maximum,
directly from the magnon spectral functions (Fig.~\ref{fig3})
corresponding to the first non-zero $q$ values from different system
sizes. The extracted linewidths are plotted as a function of the
wave-vector in Figs.~\ref{fig6a} and ~\ref{fig6b}, for $x=0.03$ and
0.05, respectively. In order to have sufficiently small $q$ values,
and also check for the probable finite-size effects we have performed
the calculations on system sizes ranging from $200a\times200a$ up to
$500a\times500a$. The linewidth data are averaged over one hundred
disorder configurations and the error bars corresponding to the
standard deviation are contained within the symbols.  Now, since we
are interested in the $q\rightarrow 0$ regime, we focus on a
restricted region of the $q$ values, (highlighted by the shaded
regions in the plots), in order to give more weight to the smallest
available $q$'s. We remark that the limit considered for the shaded
regions only serve as an approximate value and not as a clear
demarcation of the $q$ regime, defining the long wavelength
limit. Note that the value of $\ln(qa)\approx -4$ corresponds to a
value $qa\approx 0.02$.  To determine the $q$-dependence we use a
linear fit of the form $n\ln(qa)+C$, (with $n=3$, 4, and 5) for the
data within these shaded regions.  As can be clearly seen for both
cases, $x=0.03$ and 0.05, it is the $n=4$ fit (denoted by the solid
line) which best describes the linewidth behavior in this region.
Beyond this region, the linewidth begins to deviate from this behavior
although the deviations are less for $x=0.05$ compared to $x=0.03$.
Also note that the same $q$-scaling was observed for the other
concentrations as well. This clearly shows that in the long wavelength
limit and at low temperatures the intrinsic linewidth actually scales
as $q^4$ in these two-dimensional systems. 
Our findings are interestingly in good agreement with the prediction of a 
$q^{d+2}$- dependence reported in Ref.~\onlinecite{christou1986}. This agreement is not obvious since in the 
latter work a different analytical approach, based on diagrammatic perturbation theory, was used and also 
the couplings were restricted to nearest neighbors only. Whereas our study is more general in the sense 
that the couplings are extended, as well as the linewidth is extracted directly from the magnon spectral functions. 
In this context, it is worth mentioning that studies based on virtual-crystal-like approaches often lead to an infinite 
lifetime, since the spin fluctuations are unaccounted and the disorder effects are neglected, implying no mechanism for  
magnon decay. However, disorder plays an essential role, as shown here, in leading to a finite lifetime in these systems. 
From this $q^4$ scaling we infer that, in the long wavelength limit, the linewidth is actually smaller than the
excitation energy which was qualitatively clear from the well-defined
peaks observed around the $\Gamma$-point in the spectral
functions. Nevertheless, as we have seen, it is difficult to identify
 precisely the values of $q$ below which this behavior holds and these
values, in turn, should also depend on the concentration $x$.
Similar difficulties were also demonstrated in the case of three-dimensional 
systems\cite{akash2015} where a $q^3$ behavior, instead of $q^5$, was observed if the 
considered $q$ values were not sufficiently small. Apparently, in the present case we 
do not observe any clear crossover from the $q^4$ scaling to any other form within 
the considered range of the wave-vectors. We also conclude that the scaling of the 
linewidths does in fact depend on the dimensionality.

%--------------------------------------------------------------------------------
%--------------------------------------------------------------------------------

\ As already mentioned, the energy and the linewidth calculated from
the moments do not coincide with the real ones extracted from the
spectral function. The reason behind is that moments can only
reproduce the characteristic features of a distribution when they are
perfectly symmetric, such as a Gaussian or a Lorentzian. In the
present case, the spectral functions are actually asymmetric and hence
the moments prove to be inappropriate to estimate the real line shape
and the peak positions. In the clean case (absence of any disorder)
one is likely to get reliable results from the moments analysis of the
spectral functions as the excitations can be completely
symmetric. However, disorder leads to a strong asymmetry of the
excitations, as we have seen in the present case. There is a
considerable broadening in the spectrum observed especially close to
the zone boundary. Thermal fluctuations also play an important role in
these systems, but since we focus only on the low-temperature
excitations we can neglect the thermal effects here. Further
experimental studies to quantitatively examine the linewidth in these
compounds could prove to be very useful.
%--------------------------------------------------------------------------------
%--------------------------------------------------------------------------------
\section{CONCLUSION}
\label{sect:Conclusions}
%--------------------------------------------------------------------------------
%--------------------------------------------------------------------------------
\ We have addressed the low temperature spin excitations in
two-dimensional diluted Heisenberg systems, with a particular focus on
the long wavelength limit. A self-consistent Green's functions based
approach is used to evaluate the magnon DOS and the dynamical spectral
functions. Well-defined excitations are observed only in a restricted
region of the Brillouin zone, around the $\Gamma$-point. It is
demonstrated that determining the correct wave-vector dependence of
the magnon linewidth in diluted systems is not an ordinary 
task. Contrary to some previous studies, we have shown that the
moments associated with the spectral function are inappropriate to
determine the linewidth or the excitation energies. The moments
overestimate the real spin stiffness as well as provide a linear
$q$-dependence of the linewidth, implying incoherent excitations in
the limit $q\rightarrow 0$. However, this is found to be inconsistent
with the stiffness and the linewidth extracted from the calculated
spectral functions. In the long wavelength limit, the linewidth in
fact scales as $q^4$ in two-dimensional systems, for a wide range of
impurity concentrations. The discrepancy arises due to the inability
of the moments to reproduce the asymmetry in the excitation peaks. The
origin of this asymmetry, and thus a finite lifetime, is ascribed to the disorder induced
broadening of the spin waves. Hence, this underlines the importance of
the disorder effects in these systems and we emphasize that the
failure to properly account for them will certainly result in an
incorrect wave vector dependence of the linewidth.
%--------------------------------------------------------------------------------

\ Most data storage devices, in nowadays spintronics, try to manipulate 
the dynamical motion of spins. From this perspective, a precise
knowledge of the excitations' lifetime (inversely proportional to the
linewidth) could be of practical relevance. For example, a short
lifetime is important for memory devices to leave a bit in a steady
state after a read-in or read-out operation. On the contrary, a
longer lifetime is advantageous for the unhampered transmission of
signals in inter-chip communications. It would be equally interesting
to look into the temperature effects on the spin dynamics, in
particular the linewidth, where in addition to disorder the thermal
effects also play a vital role. However, this is beyond the scope of
the current work. The present findings provide qualitative insights
into the low temperature excitations and the magnon lifetimes in
two-dimensional ferromagnets, and could serve as a firm basis for
future research on complex disordered magnets. More experimental
studies oriented in this direction are also highly desirable to
resolve the controversy arising from the numerous theoretical
proposals.
%--------------------------------------------------------------------------------
%--------------------------------------------------------------------------------
\acknowledgments 
\ We acknowledge financial support by DFG within the
collaborative research center SFB 689. AC would like to thank Georges Bouzerar 
for insightful comments and discussions.
%--------------------------------------------------------------------------------
\appendix
\section{Impurity Configuration Energies}
\label{append}
%--------------------------------------------------------------------------------
%--------------------------------------------------------------------------------
\begin{figure}[ht]
\centering
  \includegraphics[width=0.7\columnwidth]{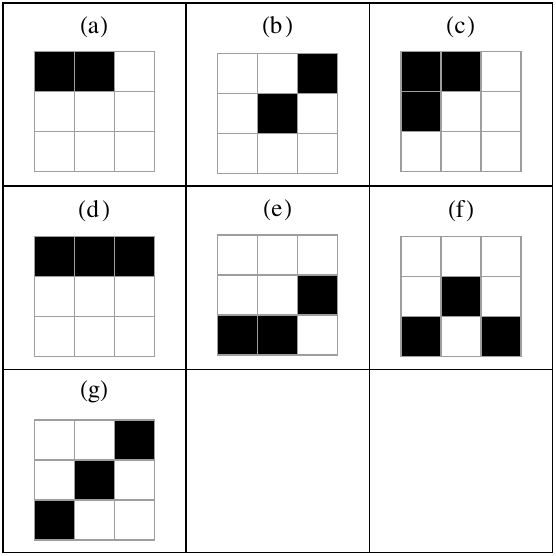}
  \caption{Basic impurity configurations on a square lattice which
    give a high contribution to the magnon DOS.}
  \label{fig7}
\end{figure}
%--------------------------------------------------------------------------------
The peaks in the averaged magnon DOS in the diluted case,
Fig.~\ref{fig1}, can be related to different configurations of some
few impurities in coordinate space on the square lattice. This can be
motivated by considering the distribution of the local magnon DOS in
Fig.~\ref{fig2a},\ref{fig2b} which reveals clusters of less than four
impurities up to an energy of $\omega/(J_0S^2)\approx 3$. Thus, the
relevant small clusters which give rise to the magnon DOS peaks on a
square lattice are identified to be those plotted in Fig.~\ref{fig7}.
The energies corresponding to the configurations (a)-(g) are given by
\allowdisplaybreaks
\begin{align}
  E_{\text{(a)};1} ={}& 2\label{simple_imp_config1}\\
  E_{\text{(b)};1} ={}& 2^{-\alpha/2}E_{\text{(a)};1}\\
  E_{\text{(c)};1} ={}& 3\\
  E_{\text{(c)};2} ={}& 1+2^{1-(\alpha/2)}\\
  E_{\text{(d)};1} ={}& 3\\
  E_{\text{(d)};2} ={}& 1+2^{1-\alpha}\\
  E_{\text{(e)};1/2} ={}& 1+2^{-\alpha/2}+5^{-\alpha/2}\pm 10^{-\alpha/2}[2^\alpha\nonumber\\
  -10^{\alpha/2}{}&(1+2^{\alpha/2})+5^\alpha(1-2^{\alpha/2}+2^\alpha)]^{
    \frac{1}{2}}\\
  E_{\text{(f)};1}  ={}& 2^{-\alpha/2}E_{\text{(c)};1}\\
  E_{\text{(f)};2}  ={}& 2^{-\alpha/2}E_{\text{(c)};2}\\
  E_{\text{(g)};1}  ={}& 2^{-\alpha/2}E_{\text{(d)};1}\\
  E_{\text{(g)};2}  ={}& 2^{-\alpha/2}E_{\text{(d)};2}\label{simple_imp_config2}
\end{align}
where the energy $E_{\text{(.)};p}$ is given in values of
$J_0S^2/a^\alpha$, with $a$ being the lattice constant and the index
$p$ the eigenvalue number (the ground-state energy zero has been
excluded).
%--------------------------------------------------------------------------------
\begin{figure}[ht]
\centering
\subfigure[]{
  \includegraphics[width=\columnwidth]{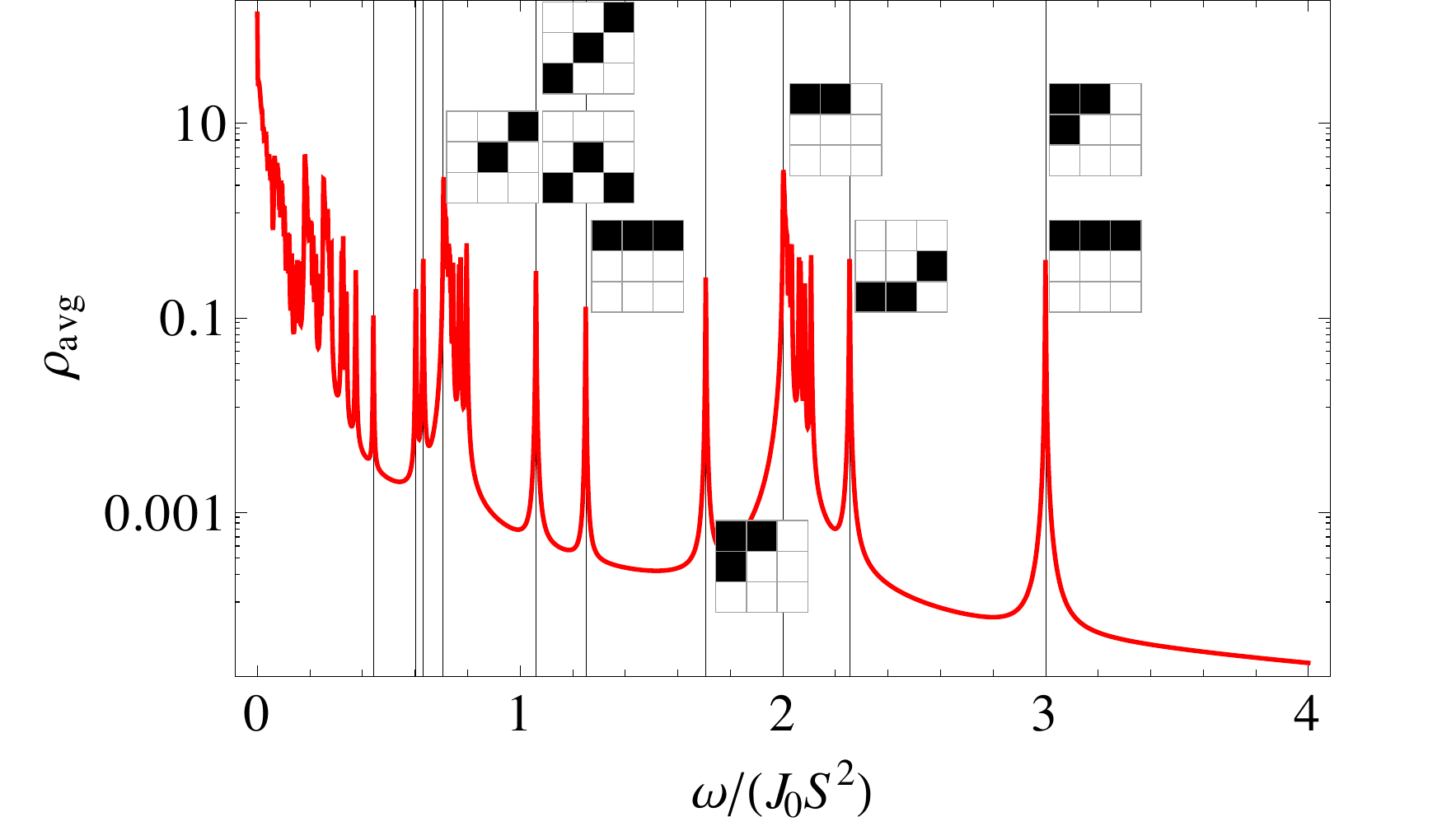}
}
\subfigure[]{
 \includegraphics[width=0.92\columnwidth]{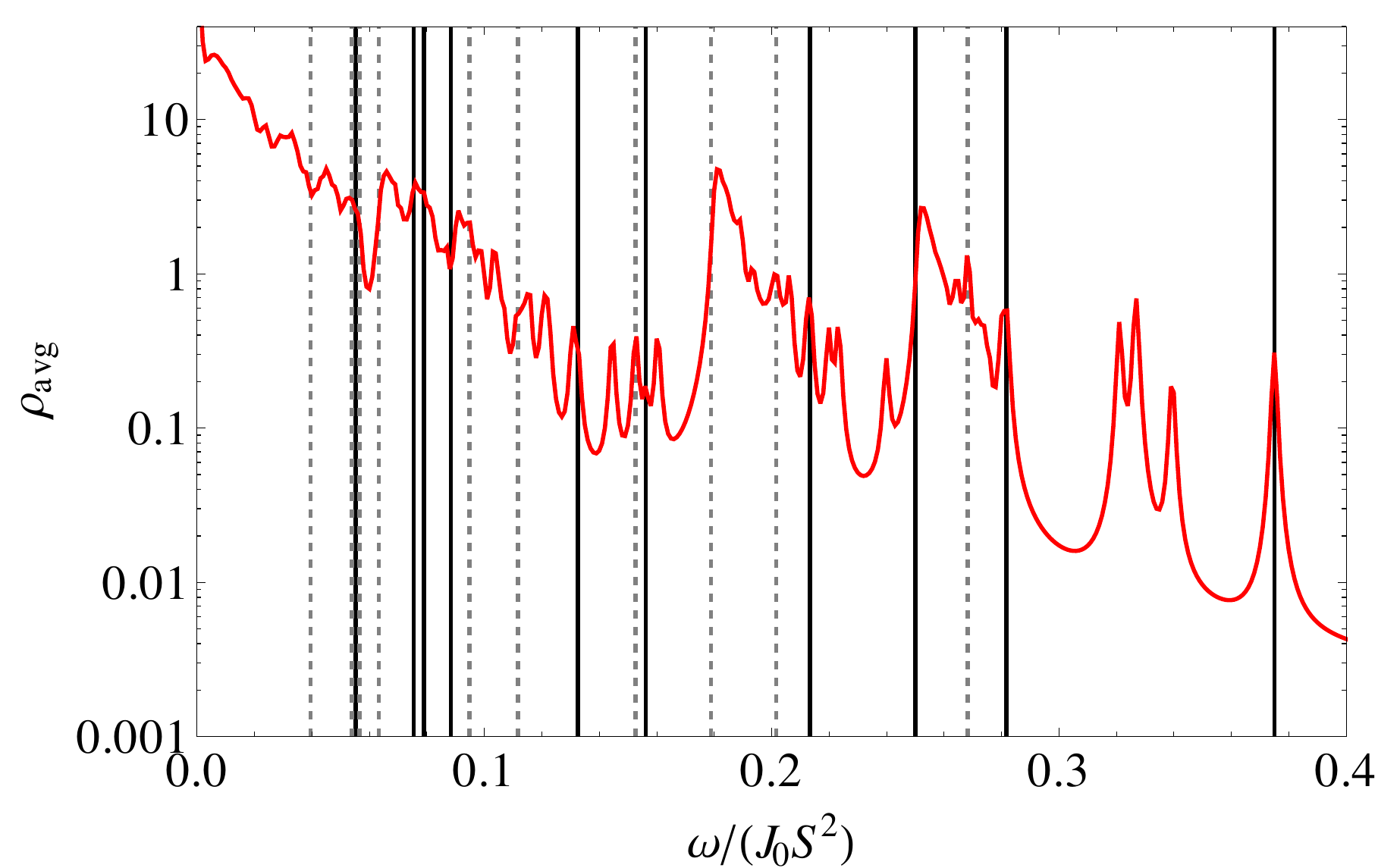}
}
\caption{(Color online) Averaged magnon DOS $\rho_{\text{avg}}$ of a small system
  $L=10a$ with only three impurities.\newline (a) For specific maxima
  we show the corresponding impurity configuration in coordinate space
  on the square lattice. The energy corresponding to a configuration
  inset is on its LHS indicated by a solid line, according to
  Eq.~\ref{simple_imp_config1}-~\ref{simple_imp_config2} with
  $\alpha=3$.\newline (b) Zoom of plot (a) at small $\omega$
  values. The solid lines indicate the energies after rescaling the
  lattice constant by two, the dashed lines a rescaling by $\sqrt{5}$.
}
  \label{fig8}
\end{figure}
%--------------------------------------------------------------------------------
The energies $E_{(.);p}$ are indicated in Fig.~\ref{fig8} by vertical
lines. Fig.~\ref{fig8} shows the magnon DOS of a $10a\times 10a$
system with three impurities averaged over all possible
configurations. A comparison of this results with the $x=0.02$ case in
Fig.~\ref{fig1} (system size of $L=1340a$) reveals indeed that the
relevant energies $E\gtrsim 0.5J_0S^2$ are given by the
Eqs.~(\ref{simple_imp_config1})-(\ref{simple_imp_config2}), with
$\alpha=3$. Many of the other configurations can be generated by a
simple isotropic rescaling of the configurations shown in
Fig.~\ref{fig7}. The corresponding energies are indicated by dashed
and solid lines in Fig.~\ref{fig8}(b).
%--------------------------------------------------------------------------------
\begin{figure}[ht]
\centering
  \includegraphics[width=0.8\columnwidth]{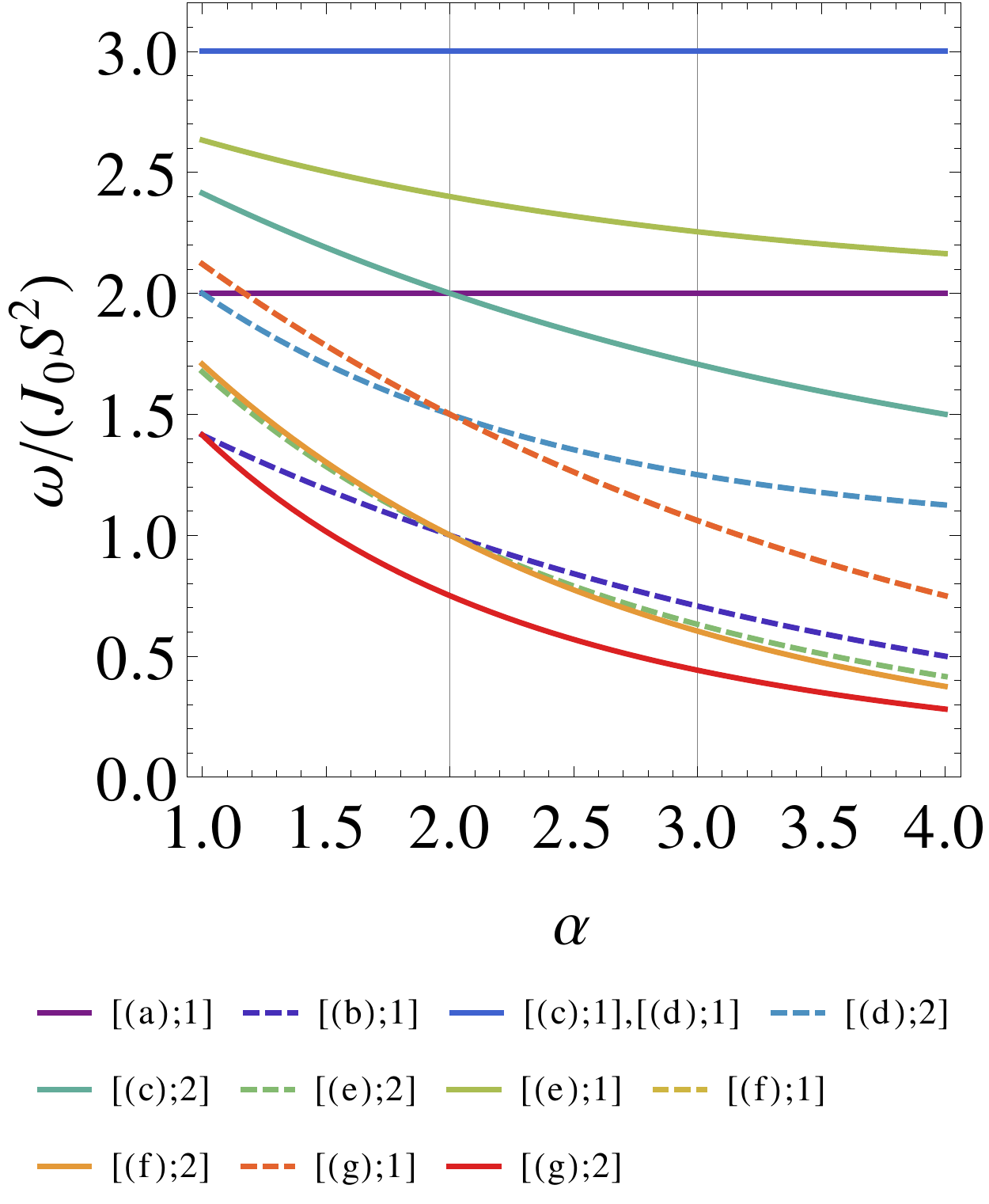}
  \caption{(Color online) Magnon energies given by
    Eq.~\ref{simple_imp_config1}-~\ref{simple_imp_config2} for the
    corresponding impurity configurations plotted in Fig.~\ref{fig8}
    as function of the exponent $\alpha$ with $\mid J_{\textbf r}\mid
    \propto r^{-\alpha}$.}
  \label{energies_alpha}
\end{figure}
%--------------------------------------------------------------------------------
To answer the question how specific the choice of the exponent
$\alpha=3$ is, we plotted the energies
Eq.~\ref{simple_imp_config1}-~\ref{simple_imp_config2} of the magnon
DOS peaks for different values of $\alpha$ in
Fig.~\ref{energies_alpha}. We recall that a two-dimensional system
with monotonically decaying interactions cannot have ferro- or
anti-ferromagnetic long-range order at $T> 0$, if $\alpha \ge 4$. The
plot shows that in the range of $2<\alpha<4$ the energies of the
individual impurity configurations do not intersect each other. Thus,
one can assume to have no significant and quantitative change in the
physics if the exponent is varied within this range.

%%%%%%%%%%%%%%%%%%%%%%%%%%%%%%%%%%
%	bibliography		 %
%%%%%%%%%%%%%%%%%%%%%%%%%%%%%%%%%%
\pagebreak

%--------------------------------------------------------------------------------
%--------------------------------------------------------------------------------
\end{document}